\documentstyle[prl,aps,multicol,graphicx]{revtex}
\newcommand{\ua}{\uparrow}
\newcommand{\da}{\downarrow}
\begin{document}

\title{An exact representation of the fermion dynamics in terms of 
Poisson processes \\ and its connection with Monte Carlo algorithms}
\author{Matteo Beccaria,$^a$ Carlo Presilla,$^b$ 
Gian Fabrizio De Angelis,$^a$ and Giovanni Jona-Lasinio$^b$}
\address{
$^a$Dipartimento di Fisica, Universit\`a di Lecce, Via Arnesano,
73100 Lecce, Italy \\
$^b$Dipartimento di Fisica, Universit\`a di Roma ``La Sapienza,''
Piazzale A. Moro 2, 00185 Roma, Italy \\
and INFM Unit\`a di Ricerca di Roma ``La Sapienza''}
\date{submitted to Physical Review Letters, October 26, 1998}
\maketitle
\begin{abstract}
We present a simple derivation of a Feynman-Kac type formula
to study fermionic systems.
In this approach the real time or the imaginary time dynamics is 
expressed in terms of the evolution of a collection of Poisson processes.
A computer implementation of this formula leads to a
family of algorithms parametrized by the values of the jump rates
of the Poisson processes.
From these an optimal algorithm can be chosen 
which coincides with the Green Function Monte Carlo 
method in the limit when the latter becomes exact.
\end{abstract}

\begin{multicols}{2}
\narrowtext

A crucial issue in quantum Monte Carlo methods is the choice
of the most convenient stochastic process to be used in the
simulation of the dynamics of the system.
This aspect is particularly evident in the case of fermion systems
\cite{TC,LINDEN,ZCG,CS}
due to the anticommutativity of the variables involved which for long 
time have not lent themselves to direct numerical evaluation.
In a recent paper \cite{DJS} progress has been made in this direction
by providing exact probabilistic expressions for quantities 
involving variables belonging to Grassmann or Clifford algebras.
In particular, the real time or the imaginary time evolution operator 
of a Fermi system or a Berezin integral can be expressed in terms 
of an associated 
stochastic dynamics of a collection of Poisson processes.
This approach depends on an older general formula to represent 
probabilistically the solution of a system of ordinary differential 
equations (ODE) in terms of Poisson 
processes \cite{DAJLS}. 
In this paper, we present a simple derivation of a similar formula 
to study fermionic systems, in particular, the Hubbard model.
However, the fermionic nature of the Hamiltonian plays
no special role and similar representations can be written down for 
any system described by a finite Hamiltonian matrix.
A computer implementation of this formula leads to a
family of algorithms parametrized by the values of the jump rates
of the Poisson processes.
For an optimal choice of these parameters we obtain an algorithm
which coincides with the well known Green Function Monte Carlo 
method in the limit when the latter becomes exact \cite{CBS}.
In this way we provide a theoretical characterization of GFMC.

Let us consider the Hubbard Hamiltonian
\begin{eqnarray}
\label{Hubbard}
H &=& - \sum_{i=1}^{|\Lambda|} \sum_{j=i+1}^{|\Lambda|} 
\sum_{\sigma=\ua\da} \eta_{ij}
(c^\dag_{i\sigma} c^{}_{j\sigma} + c^\dag_{j\sigma} c^{}_{i\sigma}) 
\nonumber \\ &&+ 
\sum_{i=1}^{|\Lambda|} \gamma_i~ 
c^\dag_{i\ua} c^{}_{i\ua}~c^\dag_{i\da} c^{}_{i\da},
\end{eqnarray}
where $\Lambda\subset Z^d$ is a finite $d$-dimensional lattice
with cardinality $|\Lambda|$,
$\{1, \dots,|\Lambda|\}$ some total ordering of the lattice points,
and $c_{i \sigma}$  the usual anticommuting destruction operators
at site $i$ and spin index $\sigma$.
In this paper, we are interested in evaluating the matrix elements
$\langle \bbox{n}' |e^{- Ht} | \bbox{n} \rangle$
where $\bbox{n}= (n_{1 \ua},n_{1 \da}, \ldots, n_{|\Lambda| \ua},
n_{|\Lambda| \da})$ are the occupation numbers taking the values 0 or 1
\cite{NOTE1}.
The total number of fermions per spin component is a conserved
quantity, therefore we consider only configurations $\bbox{n}$ and 
$\bbox{n}'$ such that 
$\sum_{i=1}^{|\Lambda|} n'_{i\sigma} = \sum_{i=1}^{|\Lambda|} n_{i\sigma}$
for $\sigma = \ua \da$.
In the following we shall use the mod 2 addition
$n \oplus n'= (n+n') \bmod 2$.

Let $\Gamma =\{(i,j), 1\le i< j\le |\Lambda|: \eta_{ij}\neq 0\}$ 
and $|\Gamma|$ its cardinality.
For simplicity, we will assume that $\eta_{ij}=\eta$ 
if $(i,j) \in \Gamma$ and $\gamma_i=\gamma$.
By introducing
\begin{eqnarray}
\label{lambda}
\lambda_{ij \sigma}(\bbox{n}) &\equiv& 
\langle \bbox{n} \oplus \bbox{1}_{i\sigma} \oplus \bbox{1}_{j\sigma}|
c^\dag_{i\sigma} c^{}_{j\sigma} + c^\dag_{j\sigma} c^{}_{i\sigma}
|\bbox{n}\rangle 
\nonumber \\ &=&
(-1)^{n_{i\sigma} + \cdots + n_{j-1 \sigma}}
\nonumber \\ &&\times 
\left[ n_{j \sigma} (n_{i \sigma} \oplus 1) -
n_{i \sigma} (n_{j \sigma} \oplus 1) \right],  
\end{eqnarray}
where $\bbox{1}_{i\sigma}=(0,\ldots,0,1_{i\sigma},0,\ldots,0)$, and 
\begin{eqnarray}
V(\bbox{n}) &\equiv& 
\langle \bbox{n}|H|\bbox{n}\rangle =
\gamma \sum_{i=1}^{|\Lambda|} n_{i \ua} n_{i \da}, 
\label{potential}
\end{eqnarray}
the following representation holds
\begin{eqnarray}
\label{TheFormulaa}
\langle \bbox{n}'|e^{-Ht} | \bbox{n}\rangle =  \bbox{E} \left(
\delta_{ \bbox{n}' , \bbox{n} \oplus \bbox{N}^t } 
{\cal M}^t \right)
\end{eqnarray}
\begin{eqnarray}
\label{TheFormulab}
{\cal M}^t &=& \exp \biggl\{ 
\sum_{(i,j)\in\Gamma} \sum_{\sigma=\ua\da} \int_{[0,t)} \!\!\!\!\!\!
\log \left[ \eta \rho^{-1} 
\lambda_{ij \sigma} (\bbox{n} \oplus \bbox{N}^s) \right] dN^s_{ij\sigma} 
\nonumber \\ && 
- \int_0^t V(\bbox{n} \oplus \bbox{N}^{s}) ds 
+ 2|\Gamma|\rho t \biggr\}.
\end{eqnarray}
Here, $\{N_{ij\sigma}^t\}$, $(i,j) \in \Gamma$,  is a family of 
$2|\Gamma|$ independent Poisson processes with parameter $\rho$ and 
$\bbox{N}^t= (N_{1 \ua}^t,N_{1 \da}^t, \ldots, N_{|\Lambda| \ua}^t,
N_{|\Lambda| \da}^t)$ are $2|\Lambda|$ stochastic processes defined as
\begin{equation}
N^t_{k \sigma} = \sum_{
(i,j) \in \Gamma:~i=k~{\rm or}~j=k }
N^t_{ij \sigma}.
\end{equation}
We remind that a Poisson process $N^t$ with parameter $\rho$ is a jump 
process characterized by the following probabilities:
\begin{equation}
P\left( N^{t+s} - N^t=k \right) = {(\rho s)^k \over k !} e^{-\rho s}.
\label{Pp}
\end{equation}
Its trajectories are piecewise-constant increasing integer-valued 
functions continuous from the left.
The stochastic integral $\int dN^t$ is just an ordinary 
Stieltjes integral
\begin{eqnarray}
\int_{[0,t)} f(s,N^s) dN^s = \sum_{k:~ s_k <t} f(s_k,N^{s_k}),
\nonumber
\end{eqnarray}
where $s_k$ are random jump times having probability density
$p(s)=\rho e^{- \rho s}$.
Finally, the symbol $\bbox{E}( \ldots )$ is the expectation of the 
stochastic functional within braces.
We emphasize that a similar representation holds for the real
time matrix elements $\langle \bbox{n}' |e^{- iHt} | \bbox{n} \rangle$.

Summarizing, we associate to each $\eta_{ij} \neq 0$ 
a link connecting the sites $i$ and $j$ and assign to it a pair of 
Poisson processes $N^t_{ij \sigma}$ with $\sigma=\ua\da$. 
Then, we assign to each site $i$ and spin component $\sigma$ 
a stochastic process $N^t_{i \sigma}$ which is the sum of all 
the processes associated with the links incoming at that site
and having the same spin component.
A jump in the link process $N^t_{ij \sigma}$ implies a jump in
both the site processes $N^t_{i \sigma}$ and $N^t_{j \sigma}$. 
Equations (\ref{TheFormulaa}) and (\ref{TheFormulab}) are immediately
generalizable to non identical parameters $\eta_{ij}$ and $\gamma_i$. 
In this case, it may be convenient to use Poisson processes 
$N^t_{ij \sigma}$ with different parameters $\rho_{ij\sigma}$. 

We now show that the representation (\ref{TheFormulaa}-\ref{TheFormulab})
follows from the general formula to represent probabilistically
the solution of an ODE system \cite{DAJLS}
and the expression of the matrix elements of $H$.
The matrix elements $\langle \bbox{n}'|e^{-Ht} | \bbox{n}\rangle$
obey the ODE system
\begin{eqnarray}
{d \over dt}
\langle \bbox{n}'|e^{-Ht} | \bbox{n}\rangle =  
- \sum_{\bbox{n}''} \langle \bbox{n}'| H | \bbox{n}''\rangle
\langle \bbox{n}''| e^{-Ht} | \bbox{n}\rangle.
\label{ODE}
\end{eqnarray}
One may check that (\ref{TheFormulaa}-\ref{TheFormulab}) is indeed
solution of (\ref{ODE}) by applying the rules of stochastic 
differentiation. 
We have
\begin{eqnarray}
&\bbox{E}& \left( 
\delta_{ \bbox{n}' , \bbox{n} \oplus \bbox{N}^{t+dt}}
{\cal M}^{t+dt} \right)
\nonumber \\ &&= 
\sum_{\bbox{n}''} \bbox{E} \biggl( 
\prod_{(i,j)\in\Gamma} \prod_{\sigma=\ua\da}
\delta_{ \bbox{n}' , \bbox{n}'' \oplus d \bbox{N}^t}
\nonumber \\ &&~~\times
e^{ \int_{[t,t+dt)}
\log \left[ \eta \rho^{-1} \lambda_{ij \sigma} 
(\bbox{n} \oplus \bbox{N}^s) \right] dN^s_{ij\sigma} }  
\nonumber \\ &&~~\times
e^{- V(\bbox{n} \oplus \bbox{N}^{t}) dt 
+ 2|\Gamma|\rho dt } 
~\delta_{ \bbox{n}'' , \bbox{n} \oplus \bbox{N}^t} {\cal M}^t \biggr)
\label{dM}
\end{eqnarray}
For the Markov property, the expectation of the factors 
containing the stochastic integrals in the interval $[t,t+dt]$ can
be taken separately. 
By expanding each one of them over all possible numbers of
jumps of the Poisson processes as
\begin{eqnarray}
&\bbox{E} &\left( \delta_{ \bbox{n}' , \bbox{n}'' \oplus d \bbox{N}^t}
~e^{\int_{[t,t+dt)} 
\log \left[ \eta \rho^{-1} \lambda_{ij \sigma} 
(\bbox{n} \oplus \bbox{N}^s) \right] dN^s_{ij\sigma} } \right) 
\nonumber \\ &&=
\delta_{ \bbox{n}' , \bbox{n}'' } ~e^0~e^{-\rho dt} 
\nonumber \\&&~~+ 
\delta_{ \bbox{n}' , \bbox{n}'' \oplus \bbox{1}_{i\sigma}
\oplus \bbox{1}_{j\sigma} }
~e^{\log \left[ \eta \rho^{-1} \lambda_{ij \sigma} 
(\bbox{n} \oplus \bbox{N}^t) \right] } 
~e^{-\rho dt} \rho dt + \ldots
\nonumber \\ &&=
\delta_{ \bbox{n}' , \bbox{n}'' } + 
\left[ \delta_{ \bbox{n}' , \bbox{n}'' \oplus \bbox{1}_{i\sigma}
\oplus \bbox{1}_{j\sigma} }
\eta \rho^{-1} \lambda_{ij \sigma} (\bbox{n} \oplus \bbox{N}^t) \right.
\nonumber \\&&~~
\left. - \delta_{ \bbox{n}' , \bbox{n}'' } \right] \rho dt 
+ {\cal O}\left( dt^2 \right),
\nonumber
\end{eqnarray}
up to order $dt$ we obtain
\begin{eqnarray}
&\bbox{E}& \left( 
\delta_{ \bbox{n}' , \bbox{n} \oplus \bbox{N}^{t+dt}}
{\cal M}^{t+dt} \right) 
\nonumber \\ &&=
\sum_{\bbox{n}''} \biggl[ \delta_{ \bbox{n}' , \bbox{n}''} 
+ \sum_{(i,j)\in\Gamma} \sum_{\sigma=\ua\da} 
\delta_{ \bbox{n}' , \bbox{n}'' \oplus \bbox{1}_{i\sigma}
\oplus \bbox{1}_{j\sigma} }
\eta \lambda_{ij \sigma} (\bbox{n}'') dt
\nonumber \\ &&~~- 
\delta_{ \bbox{n}' , \bbox{n}''} V(\bbox{n}'') dt \biggr] 
\bbox{E} \left(
\delta_{ \bbox{n}'' , \bbox{n} \oplus \bbox{N}^t}
 {\cal M}^t \right).
\end{eqnarray}  
Finally, we rewrite this relation as
\begin{eqnarray}
&d \bbox{E}& \left( 
\delta_{ \bbox{n}' , \bbox{n} \oplus \bbox{N}^t}
{\cal M}^t \right) 
\nonumber \\&&= 
\bbox{E} \left( 
\delta_{ \bbox{n}' , \bbox{n} \oplus \bbox{N}^{t+dt}}
{\cal M}^{t+dt} \right) 
- \bbox{E} \left( 
\delta_{ \bbox{n}' , \bbox{n} \oplus \bbox{N}^t}
{\cal M}^t \right) 
\nonumber \\ &&=
- \sum_{\bbox{n}''} 
\langle \bbox{n}'| H | \bbox{n}'' \rangle
\bbox{E} \left( \delta_{ \bbox{n}'' , \bbox{n} \oplus \bbox{N}^t } 
 {\cal M}^t  \right) dt. 
\label{final}
\end{eqnarray}
It is clear that the fermionic nature of the Hamiltonian $H$ plays
no special role in the above derivation.
For later use, note that summing (\ref{final}) over $\bbox{n}'$
we have
$d \bbox{E}({\cal M}^t) = -\bbox{E} ({\cal H}^t  {\cal M}^t) dt$,
where
\begin{eqnarray}
{\cal H}^t &\equiv& \sum_{\bbox{n}'} 
\langle \bbox{n}'| H | \bbox{n} \oplus \bbox{N}^t \rangle
\nonumber \\ &=&
- \eta \sum_{(i,j)\in\Gamma} \sum_{\sigma=\ua\da}
 \lambda_{ij \sigma} (\bbox{n} \oplus \bbox{N}^t)
+ V(\bbox{n} \oplus \bbox{N}^{t}). 
\label{H}
\end{eqnarray}

In order to construct an efficient algorithm for evaluating 
(\ref{TheFormulaa}-\ref{TheFormulab}), 
we start by observing that the functions 
$\lambda_{ij\sigma}(\bbox{n} \oplus \bbox{N}^s)$ vanish 
when the occupation numbers $n_{i\sigma} \oplus N_{i\sigma}^s$ and 
$n_{j\sigma} \oplus N_{j\sigma}^s$ are equal.
We say that for a given value of $\sigma$ the link $ij$ is active
at time $s$ if $\lambda_{ij\sigma}(\bbox{n} \oplus \bbox{N}^s)\neq 0$.
We shall see in a moment that only active links are relevant. 
Let us consider how the stochastic integral in (\ref{TheFormulab}) builds
up along a trajectory defined by considering the time ordered 
succession of jumps in the family $\{ N^t_{ij\sigma} \}$.
The contribution to the stochastic integral in the 
exponent of (\ref{TheFormulab}) at the first jump time of a link, 
for definiteness suppose that the link $i_1j_1$ with spin component
$\sigma_1$ jumps first at time $s_1$, is
\begin{eqnarray}
\log \left[ \eta \rho^{-1} \lambda_{i_1j_1 \sigma_1} 
(\bbox{n} \oplus \bbox{N}^{s_1}) \right] ~\theta(t-s_1),
\nonumber
\end{eqnarray}
where $\bbox{N}^{s_1}=\bbox{0}$ due the assumed left continuity.
Therefore, if the link $i_1j_1\sigma_1$ was active at time 0 we obtain 
a finite
contribution to the stochastic integral otherwise we obtain $-\infty$.
If $s_1 \geq t$ we have no contribution to the stochastic 
integral from this trajectory. 
If $s_1 < t$ a second jump of a link, suppose $i_2j_2$ with spin
component $\sigma_2$, can take place
at time $s_2>s_1$ and we obtain a contribution 
\begin{eqnarray}
\log \left[ \eta \rho^{-1} \lambda_{i_2j_2 \sigma_2} 
(\bbox{n} \oplus \bbox{N}^{s_2}) \right] ~\theta(t-s_2).
\nonumber
\end{eqnarray}
The analysis can be repeated by considering an arbitrary number of jumps.
Of course, when the stochastic integral is $- \infty$, which is the 
case when some $\lambda=0$, there is no contribution to the expectation. 
The other integral in (\ref{TheFormulab}) is an ordinary integral of
a piecewise constant bounded function. 

We now describe the algorithm. 
From the above remarks it is clear that the only trajectories to be 
considered are those associated to the jumps of active links.
It can be seen that this corresponds to the conservation of the total 
number of fermions per spin component.
The active links can be determined after each jump by inspecting the
occupation numbers of the sites in the set $\Gamma$ according to the 
rule that the link $ij$ is active for the spin component $\sigma$ if 
$n_{i\sigma} + n_{j \sigma} =1$. 
We start by determining the active links in the initial configuration 
$\bbox{n}$ assigned at time $0$ and make an extraction with uniform 
distribution to decide which of them jumps first. 
We then extract the jump time $s_1$ according to the probability
density $p_{A_1}(s)=A_1 \rho \exp(- A_1 \rho s)$ where $A_1$ 
is the number of active links before the first jump takes place.
The contribution to ${\cal M}^t$ at the time of the first jump 
is therefore, up to the factor 
$\exp \left( -2|\Gamma|\rho t \right)$,
\begin{eqnarray}
&& \eta \rho^{-1} \lambda_{i_1j_1 \sigma_1} (\bbox{n} \oplus 
\bbox{N}^{s_1})
e^{ -V(\bbox{n} \oplus \bbox{N}^{s_1}) s_1 }
e^{-(2|\Gamma|-A_{1}) \rho s_1} 
\nonumber \\ && \times \theta(t-s_1)
+ e^{ -V(\bbox{n} \oplus \bbox{N}^{t}) t }
e^{-(2|\Gamma|-A_{1})\rho t} ~\theta(s_1-t),  
\nonumber
\end{eqnarray}
where $\exp [-(2|\Gamma|-A_{1})\rho s]$ is the probability 
that the $2|\Gamma|-A_{1}$ non active links do not jump in the 
time interval $s$.
The contribution of a given trajectory is obtained by multiplying the
factors corresponding to the different jumps until the last jump 
takes place later than $t$.
For a given trajectory we thus have 
\begin{eqnarray}
{\cal M}^t = \prod_{k \geq 1} &\Bigl [&
\eta \rho^{-1} \lambda_{i_kj_k \sigma_k} (\bbox{n} \oplus \bbox{N}^{s_k})
\nonumber \\ && \times
e^{ [A_{k} \rho -V(\bbox{n} \oplus \bbox{N}^{s_k})] (s_k-s_{k-1}) }
~\theta(t-s_k) 
\nonumber \\ && +
e^{ [A_{k} \rho -V(\bbox{n} \oplus \bbox{N}^{t})] (t-s_{k-1}) }
~\theta(s_k-t) \Bigr].
\label{Mt}
\end{eqnarray}
Here, $A_k=A(\bbox{n} \oplus \bbox{N}^{s_k})$ is the number of active 
links in the interval $(s_{k-1}, s_k]$ and $s_0=0$.
Note that the exponentially increasing factor 
$\exp \left( 2|\Gamma|\rho t \right)$ in (\ref{TheFormulab}) cancels
out in the final expression of ${\cal M}^t$.
The analogous expression of ${\cal M}^t$ for real times is simply 
obtained by replacing $\eta \to i \eta$ and $\gamma \to i \gamma$.

Let us specialize the algorithm for the calculation of the 
ground state energy $E_0$.
This can be related to the matrix elements (\ref{TheFormulaa}) in the 
following way
\begin{equation}
E_0 = -\lim_{t \to \infty} {  
\sum_{\bbox{n}'} \partial_t  \langle \bbox{n}'|e^{-Ht} | \bbox{n}\rangle 
\over 
\sum_{\bbox{n}'} \langle \bbox{n}'|e^{-Ht} | \bbox{n}\rangle }. 
\label{E0}
\end{equation}
The denominator in this expression corresponds to evaluating the 
expectation in (\ref{TheFormulaa}) without the $\delta$ and is estimated
by 
\begin{equation}
\sum_{\bbox{n}'} \langle \bbox{n}'|e^{-Ht} | \bbox{n}\rangle 
\simeq {1 \over K} \sum_{p=1}^K {\cal M}^t_p,
\label{DEN}
\end{equation}
where the index $p$ denotes one of the $K$ simulated trajectories and
${\cal M}^t_p$ is the value of ${\cal M}^t$ for the $p$th
trajectory.
The numerator of (\ref{E0}) is estimated similarly to (\ref{DEN})
with ${\cal M}^t_p$ replaced by ${\cal H}^t_p {\cal M}^t_p$, 
where ${\cal H}^t_p$ is the value for the $p$th trajectory of
${\cal H}^t$ given by (\ref{H}). 
 
The variance of the stochastic process ${\cal M}^t$ 
increases with $t$ and its distribution is not 
well estimated if the number $K$ of trajectories remains fixed.
As an alternative to increasing $K$, one may resort to
the reconfiguration method \cite{HETHERINGTON}.
A simulation with $K$ trajectories is performed for a time 
$t$ but is repeated $R$ times 
choosing randomly the initial configurations among those
reached in the previous simulation \cite{CS,S}.

In principle, the algorithms parametrized by $\rho$ are all equivalent
as (\ref{TheFormulaa}-\ref{TheFormulab}) holds for any choice of
the Poisson rates.
However, since we estimate the expectation values with a finite
number of trajectories, this may introduce a systematic error.
In Fig. 1 we show the dependence of the error 
$E_0-E_0^{\rm exact}$ as a function of  
$\rho$ in the case of a small
one-dimensional system which can be exactly diagonalized.
It is evident that the best performance of the algorithm is in 
correspondence of the natural choice $\rho \sim \eta$ 
independently of the interaction strength.
This behavior can be understood on the basis of the following 
qualitative argument.
The average number of configuration changes in the time $t$ is 
$A \rho t$ where $A$ is the average number of active links.
This is also the average number of terms in the product of (\ref{Mt}).
In absence of sign problem as in the case of Fig. 1, 
we roughly estimate the r.h.s. of (\ref{DEN}) as
$\left(\eta \rho^{-1} \right)^{A \rho t} e^{(A\rho-V)t}$,
where $V$ is the average potential.
We see that the derivative with respect to $\rho$ of this expression
vanishes for $\rho=\eta$.
It is easily seen that this corresponds also to a minimum of the
variance of ${\cal M}^t$ and, therefore, to a minimum of the 
statistical error.

The convergence features of the algorithm can be improved also
using the importance sampling method \cite{CK}.
Consider the operator $\widetilde{H}$ isospectral to $H$
defined by the matrix elements 
$\langle \bbox{n}'|\widetilde{H}|\bbox{n}\rangle =
\langle \bbox{n}'|g \rangle
\langle \bbox{n}'|H|\bbox{n}\rangle
\langle \bbox{n}|g \rangle^{-1}$, where $|g \rangle$ is a given state.
The stochastic representation (\ref{TheFormulaa}-\ref{TheFormulab}) 
and the corresponding algorithmic implementation hold unchanged for 
the new operator $\widetilde{H}$ with the substitution
$\lambda_{ij \sigma}(\bbox{n}) \to \lambda_{ij \sigma}(\bbox{n})
\langle \bbox{n} \oplus \bbox{1}_{i\sigma} \oplus \bbox{1}_{j\sigma}
|g \rangle \langle \bbox{n}|g \rangle^{-1}$.
In this way, however, the value of $\rho$ is not tuned to obtain
the best performance.
For this purpose we have to choose $\rho$ dependent on 
the indices $ij\sigma$ and on the configuration $\bbox{n}$
so that
$\rho_{ij\sigma}(\bbox{n}) =
\eta \left| 
\langle \bbox{n} \oplus \bbox{1}_{i\sigma} \oplus \bbox{1}_{j\sigma}
|g \rangle \langle \bbox{n}|g \rangle^{-1} \right|$.

We consider now the connection of our approach with the GFMC method.
In the GFMC, the ground state of the system is filtered out from
a given initial state $|\bbox{n} \rangle$ by iteratively applying
the operator $G=1-H\Omega^{-1}$.
In the occupation number representation we can write 
\begin{equation}
\langle \bbox{n}' |G| \bbox{n} \rangle =
{\cal P}_{\bbox{n}' \bbox{n}}~ b_{\bbox{n}'\bbox{n}},
\label{G}
\end{equation}
where
\begin{equation}
{\cal P}_{\bbox{n}' \bbox{n}} =
{|\langle \bbox{n}' |G| \bbox{n} \rangle| \over 
\sum_{\bbox{n}''} |\langle \bbox{n}'' |G| \bbox{n} \rangle|}
\label{P}
\end{equation}
is a stochastic matrix and
\begin{equation}
b_{\bbox{n}'\bbox{n}} =
{\langle \bbox{n}' |G| \bbox{n} \rangle \over 
|\langle \bbox{n}' |G| \bbox{n} \rangle|}
\sum_{\bbox{n}''} |\langle \bbox{n}'' |G| \bbox{n} \rangle|
\label{B}
\end{equation}
a, possibly negative, weight factor.
A trajectory is defined as a series of steps, each one of duration 
$\Omega^{-1}$, in which the configuration changes from $\bbox{n}$ to 
$\bbox{n}'$ according to the stochastic matrix 
${\cal P}_{\bbox{n}' \bbox{n}}$.
The weight of a trajectory is 
$\prod_{ \{ \bbox{n} \} } b_{\bbox{n}'\bbox{n}}$, 
where $\bbox{n} $ runs over
the intermediate configurations visited by the trajectory \cite{CS}.
Under our hypothesis $\eta_{ij}=\eta$ if $(i,j) \in \Gamma$,
the GFMC algorithm becomes very simple because all the active links
have the same probability to jump.

For large values of $\Omega$ the jump probability vanishes and 
the GFMC as described above becomes inefficient. 
As remarked in \cite{TC}, one can cope with this difficulty 
in the following way.
The probability that a configuration $\bbox{n}$ changes
at the step $n_s+1$, being unchanged in the previous $n_s$ steps, is
$p(n_s)= ({\cal P}_{\bbox{n} \bbox{n}})^{n_s} 
(1-{\cal P}_{\bbox{n} \bbox{n}})$.
Therefore, we can directly assign the weight
\begin{eqnarray}
(b_{\bbox{n}\bbox{n}})^{n_s} b_{\bbox{n}'\bbox{n}}=
\lambda_{ij\sigma} (\bbox{n})
\left\{ 1 + \left[ \eta A(\bbox{n}) - V(\bbox{n}) \right] \Omega^{-1} 
\right\}^{n_s+1}
\nonumber
\end{eqnarray}
to a configuration change $\bbox{n} \to \bbox{n}'=
\bbox{n} \oplus \bbox{1}_{i\sigma} \oplus \bbox{1}_{j\sigma}$ in which 
a fermion with spin $\sigma$ is moved from the site $i$ to the site $j$
or {\em viceversa}.
As before, $A(\bbox{n})$ is the number of active links in the 
configuration $\bbox{n}$ and $\lambda_{ij\sigma} (\bbox{n})$ and 
$V(\bbox{n})$ are given by (\ref{lambda}) and (\ref{potential}), 
respectively.
The random integer $n_s$ is extracted with probability $p(n_s)$, e.g. 
$n_s= \lfloor \log r / \log {\cal P}_{\bbox{n} \bbox{n}} \rfloor$,
where $r$ is a random number with uniform distribution in $[0,1]$.
Note that the distribution of $n_s$ becomes Poissonian with parameter
$- \log {\cal P}_{\bbox{n} \bbox{n}} \simeq \eta A(\bbox{n})\Omega^{-1}$ 
for $\Omega$ large.  
At this level, there is no apparent connection between this Poisson
process and those entering in our formula (\ref{TheFormulab}).
However, as suggested in \cite{CBS} one can take the continuum limit 
$\Omega^{-1} \to 0$ and reconstruct in this way the semigroup 
$\exp(-H t)$.
It is easy to verify that in this limit the GFMC algorithm coincides 
with that obtained from our formula for the optimal choice $\rho=\eta$.
In fact, the time interval before a configuration change takes place is 
$s=n_s \Omega^{-1}$.
For $\Omega^{-1} \to 0$, the random variable $s$ becomes continuously 
distributed in $[0, \infty)$ with probability density 
$p(s)=\Omega p(n_s)= \eta A(\bbox{n}) \exp[-\eta A(\bbox{n}) s]$
and the trajectory weight corresponding to the interval $s$ reduces to
$\lambda_{ij\sigma} (\bbox{n}) \exp \left\{ \left[ 
 \eta A(\bbox{n}) - V(\bbox{n}) \right] s \right\}$.
The coincidence holds also when importance sampling is included. 
If we change in Eqs. (\ref{G}-\ref{B}) $\Omega$ into $-i\Omega$ 
and interpret the absolute value as the modulus, in the continuum limit 
$\Omega^{-1} \to 0$ the extended GFMC algorithm coincides with 
our algorithm also for real times.

Partial support of INFN, Iniziativa Specifica RM42, is acknowledged.

\begin{figure}
\centerline{\includegraphics*[width=6.3cm,angle=90]{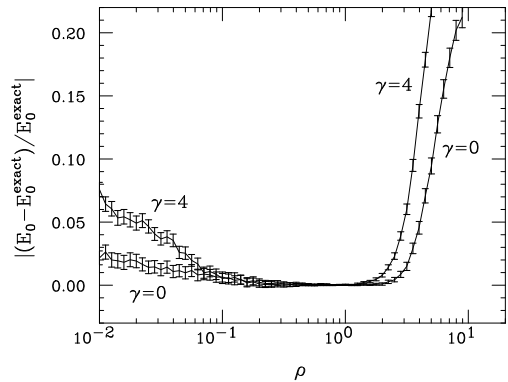}}
\caption{Absolute value of the relative error 
$\left( E_0-E_0^{\rm exact}\right) / E_0^{\rm exact}$ 
on the ground state energy of an interacting ($\gamma=4$) and 
noninteracting ($\gamma=0$) Hubbard system obtained with the present 
algorithm as a function of the Poisson parameter $\rho$. 
The system is one-dimensional with periodic boundary conditions, 
$|\Lambda|=8$,
$\sum_{i=1}^{|\Lambda|} n_{i\ua} = \sum_{i=1}^{|\Lambda|} n_{i\da}=3$, 
$\eta_{ij}= 1 \times \delta_{ij-1}$, and $\gamma_i=4$ or $\gamma_i=0$.
In the simulation we used $K=10^3$, $R=10^4$ and $t=0.26$.   
Error bars represent statistical errors.}
\end{figure}\noindent

\end{multicols}

\begin{references}
\bibitem{NOTE1} We  refer to $t$ as the physical  
time even if it is, properly speaking, an imaginary physical time.
\end{references}
\end{document}